\documentstyle[12pt]{article}
\newcommand{\be}{\begin{eqnarray}}
\newcommand{\ee}{\end{eqnarray}}

\def\lsim{\mathrel{\rlap{\lower4pt\hbox{\hskip1pt$\sim$}}
    \raise1pt\hbox{$<$}}}               
\def\gsim{\mathrel{\rlap{\lower4pt\hbox{\hskip1pt$\sim$}}
    \raise1pt\hbox{$>$}}}               

\input{epsfig.sty}

\begin{document}

\begin{center}

\rightline{}

\vspace{2cm}

\LARGE{Longitudinal and transverse structure functions\\ of proton and
deuteron at large $x$\footnote{To appear in the Proceedings of the
International Workshop on {\em JLab: Physics and Instrumentation with
6-12 GeV Beams}, Jefferson Lab (Newport News, USA), June 15-18, 1998.}}\\

\vspace{1cm}

\large{G. Ricco$^{(a)}$ and S. Simula$^{(b)}$}\\

\vspace{0.5cm}

\normalsize{$^{(a)}$Physics Department, University of Genova and INFN,
Sezione di Genova,\\ Via Dodecanneso 33, I-16146 Genova, Italy\\
$^{(b)}$Istituto Nazionale di Fisica Nucleare, Sezione Roma III,\\ Via
della Vasca Navale 84, I-00146 Roma, Italy}

\end{center}

\vspace{1cm}

\begin{abstract}

\noindent Higher-twist effects in the low-order moments of the
longitudinal and transverse structure functions of proton and deuteron
have been analyzed using available phenomenological fits of existing data
in the $Q^2$ range between $1$ and $20 ~ (GeV/c)^2$. Both twist-4 and
twist-6 contributions have been determined adopting the Natchmann
definition of moments, which allows to disentangle properly target-mass
effects. The extraction of the matrix elements of the relevant
twist-4 operators, describing quark-quark and quark-gluon correlations,
is carried out in case of the second moment. The need of transverse data
with better quality for $x \gsim 0.5$ and $Q^2 \lsim 10 ~ (GeV/c)^2$ as
well as more precise and systematic determinations of the $L / T$
separation make $JLab ~ @ ~ 12 ~ GeV$ a good place to improve our
understanding of the non-perturbative structure of hadrons.    

\end{abstract}

\newpage

\pagestyle{plain}

\section{INTRODUCTION}

\indent The experimental investigation of deep-inelastic lepton-nucleon
scattering has provided a wealth of information on the occurrence of
Bjorken scaling and its violations, giving a decisive support to the
rise of the parton model and its $QCD$-improved version, which properly
describe the logarithmic violations to scaling in the asymptotic
region. However, in the pre-asymptotic region the full dependence of
the nucleon response on the squared four-momentum transfer, $Q^2$, is
affected also by power-type corrections, which originate from
non-perturbative physics and can be analyzed in the framework provided
by the Operator Product Expansion ($OPE$). The logarithmic scale
dependence is therefore related to the so-called leading twist
operators, which in the parton language are one-body operators whose
matrix elements yield the contribution of the individual partons to the
nucleon response. On the contrary, power-type corrections are related to
higher-twist operators measuring the relevance of correlations among
partons \cite{SV82}.

\indent In case of unpolarized inelastic electron scattering the nucleon
response is described by two independent quantities: the transverse
$F_2(x, Q^2)$ and the longitudinal $F_L(x, Q^2)$ structure functions,
the latter being related to the ratio of the longitudinal to transverse
cross sections, $R_{L/T}(x, Q^2)$, by $F_L(x, Q^2) = F_2(x, Q^2) ~ (1 +
4M^2 x^2 / Q^2) R_{L/T}(x, Q^2) / [1 + R_{L/T}(x, Q^2)]$, where $x \equiv
Q^2 / 2M \nu$ is the Bjorken variable, $M$ is the nucleon mass and $\nu$
is the energy transfer in the nucleon rest frame. Systematic measurements
\cite{DATA} of the transverse function $F_2(x, Q^2)$ for proton and
deuteron targets have been carried out in the kinematical range $10^{-4}
\lsim x \lsim 1$ and for $Q^2$ values up to several hundreds of
$(GeV/c)^2$, while data on the ratio $R_{L/T}(x, Q^2)$ are available for
$0.002 \lsim x \lsim 0.8$ and $0.5 \lsim Q^2 (GeV/c)^2 \lsim 70$, though
they are still fluctuating and affected by large errors. Consequently,
phenomenological fits for both $F_2(x, Q^2)$ and $F_L(x, Q^2)$ are
available, but for the latter quantity the interpolation formulae greatly
suffer for very weak constraints. The analysis of the world data set
\cite{DATA} has allowed to extract the parton densities in the nucleon,
including their $QCD$-predicted logarithmic $Q^2$ evolution, as well as
to signal the presence of power-type scaling violations at {\em large}
$x$ ($\gsim 0.7$) and {\em low} $Q^2$ ($\lsim 10 ~ (GeV/c)^2$). The
analysis of these kinematical regions, where higher-twist effects are
important, represents the aim of the present contribution. A more
detailed version of our work will be available soon in \cite{RS98}.

\vspace{1cm}

\section{TWIST ANALYSIS}

\indent An important and effective tool for the theoretical investigation
of the complete $Q^2$ dependence of hadron structure functions is the
$OPE$, which leads to the well-known twist expansion for the moments of
the structure functions. In our analysis we do not use the Cornwall-Norton
definition of the moments, since target-mass corrections, i.e. terms
containing powers of $M^2 / Q^2$, would contribute. Instead of that we
will adopt the Natchmann definition \cite{NAT73}:
 \be
    \label{MT} 
    M_n^{(T)}(Q^2) & \equiv & \int_0^1 dx {\xi^{n+1} \over x^3} F_2(x,
    Q^2) {3 + 3 (n + 1) r + n (n + 2) r^2 \over (n + 2) (n + 3)} \\
    \label{ML}
    M_n^{(L)}(Q^2) & \equiv & \int_0^1 dx {\xi^{n+1} \over x^3}
    \left[F_L(x, Q^2) + {4M^2x \over Q^2} F_2(x, Q^2) {(n + 1) \xi - 2(n +
    2) x \over (n + 2) (n + 3)} \right]
 \ee
where $n \ge 2$, $r \equiv \sqrt{1 + 4M^2x^2 / Q^2}$ and $\xi \equiv 2x /
(1 + r)$ is the Natchmann variable. Using the {\em experimental} $F_2(x,
Q^2)$ and $F_L(x, Q^2)$ in Eqs. (\ref{MT}-\ref{ML}), target-mass effects
are canceled out and therefore the twist expansions of the experimental
$M_n^{(T)}(Q^2)$ and $M_n^{(L)}(Q^2)$ contain only {\em dynamical} twists,
namely
 \be
    \label{TWIST}
    M_n^{L(T)}(Q^2) = \sum_{\tau=2}^{\infty} ~ C_{n, \tau}^{L(T)}(Q^2 /
    \mu^2) ~ A_{n, \tau}^{L(T)}(\mu^2) ~ (\mu^2 / Q^2)^{\tau - 2 \over
    2}
 \ee  
where $\mu$ is the renormalization scale, $C_{n, \tau}^{L(T)}(Q^2 /
\mu^2)$ is a Wilson coefficient calculable within perturbative $QCD$ and
$A_{n, \tau}^{L(T)}(\mu^2)$ corresponds to the matrix elements of
operators of twist $\tau$ and spin $n$. In our analysis the expansion
(\ref{TWIST}) is simplified into
 \be
    \label{FIT}
    M_n^{L(T)}(Q^2) = A_n^{L(T)}(Q^2) + a_n^{(4)}[L(T)] {\mu^2 \over Q^2}
    \left( {\alpha_s(\mu^2) \over \alpha_s(Q^2)}
    \right)^{\gamma_n^{(4)}[L(T)]} \\ \nonumber
    + a_n^{(6)}[L(T)] {\mu^4 \over Q^4} \left( {\alpha_s(\mu^2) \over 
    \alpha_s(Q^2)} \right)^{\gamma_n^{(6)}[L(T)]}
 \ee
where $\alpha_s(Q^2)$ is the running coupling constant and
$A_n^{L(T)}(Q^2)$ is the leading-twist contribution, whose $Q^2$
dependence is calculated according to the $pQCD$ predictions at $NLO$. In
Eq. (\ref{FIT}) the last two terms in the r.h.s. are simplified
parametrizations of the twist-4 and twist-6 contributions, respectively,
as suggested in Ref. \cite{JU95}; the quantities $a_4^n[L(T)]$
($a_6^n[L(T)]$) and $\gamma_4^n[L(T)]$ ($\gamma_6^n[L(T)]$) represent the
effective magnitude and anomalous dimension of twist-4 (twist-6)
operators.

\vspace{1cm}

\section{MAIN RESULTS}

\indent Equation (\ref{FIT}) has been applied to the fit of the $Q^2$
behavior of the {\em experimental} moments $M_n^{L(T)}(Q^2)$ in the
range $1 \lsim Q^2 (GeV/c)^2 \lsim 20$. In order to evaluate the r.h.s.
of Eqs. (\ref{MT}-\ref{ML}) available phenomenological fits based on the
data of Ref. \cite{DATA} have been used and the elastic peak
contributions have been added, as it is required by the inclusive nature
of the $OPE$. In case of the deuteron, the nucleon elastic peak leads to
the quasi-elastic contribution to the inclusive cross section, which has
been evaluated through a convolution approach checked against available
$SLAC$ data (see \cite{RS98} for details). In case of the analysis of the
transverse moments, besides the free parameters $a_n^{(\tau)}$ and
$\gamma_n^{(\tau)}$ appearing in the twist-4 and twist-6 terms of Eq.
(\ref{FIT}), also the magnitude of the leading term is simultaneously
determined from the fit of our pseudo-data. On the contrary, the parton
densities of Ref. \cite{GRV95} have been adopted for evaluating
$A_n^L(Q^2)$ (see again \cite{RS98} for details). Thanks to the decoupling
of the singlet-quark and gluon operators at large $x$, the non-singlet
evolution of the leading twist can be safely applied for the analysis
of the moments with $n \ge 4$. In our work we have adopted a
renormalization scale $\mu = 1 ~ GeV$ as in \cite{JU95}.

\indent The main results of our analysis of the transverse moments for
both proton and deuteron targets can be summarized as follows: ~ i) the
simplified twist expansion (\ref{FIT}), containing up to the twist-6
term, is able to reproduce the $Q^2$ dependence of the transverse moments
starting from $Q^2 \simeq 1 ~ (GeV/c)^2$; ~ ii) the second moment
$M_2^T(Q^2)$ is only slightly affected by the twist-4 and almost
unaffected by the twist-6 (see Fig. 1(a)); ~ iii) on the contrary, both
twist-4 and twist-6 significantly contribute to the moments of order $n
\ge 4$ (see Fig. 1(b)), in accord with the presence of higher-twist
effects at large $x$; ~ iv) the signs of the twist-4 and twist-6
contributions turn out to be opposite.

\vspace{0.5cm}

\begin{figure}[htb]

\centering{\epsfxsize=16cm \epsfig{file=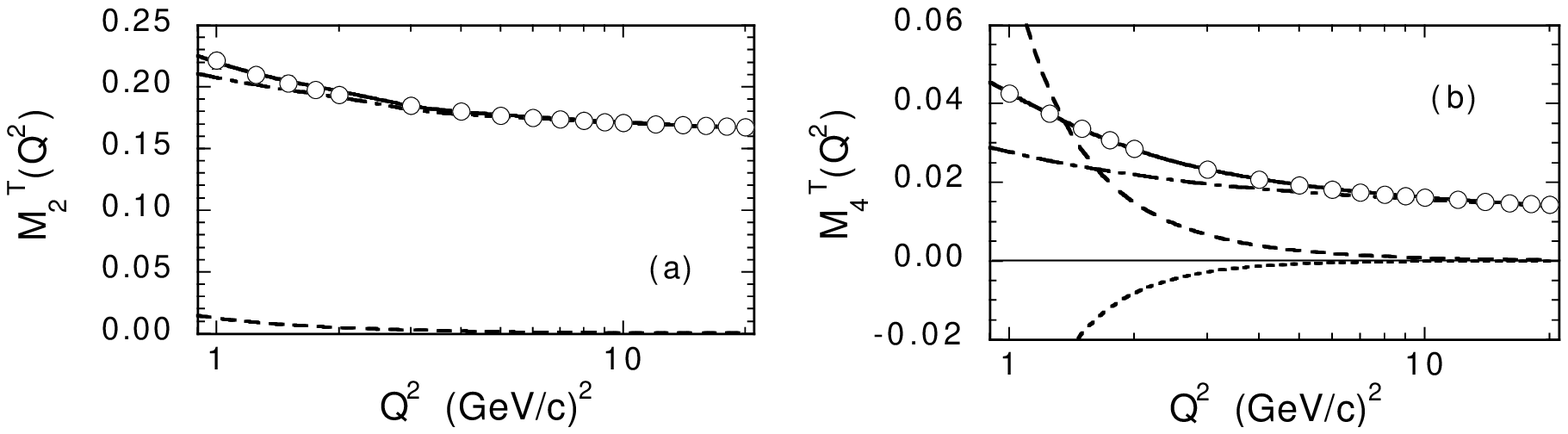}}

\vspace{0.5cm}

\parbox{0.05cm} \ $~~$ \ \parbox{15.45cm}{\small {\bf Figure 1}. Second
(a) and fourth (b) moments of the proton transverse structure function
versus $Q^2$. Open dots: pseudo-data calculated by Eq. (\ref{MT}) using
available phenomenological fits of existing data \cite{DATA} on $F_2(x,
Q^2)$; the error bars are obtained using the quoted uncertainties of the
phenomenological fits. Solid lines: result of our twist analysis based on
Eq. (\ref{FIT}); dashed-dotted, dashed and dotted lines: twist-2, twist-4
and twist-6 contributions, respectively.}

\end{figure}

\vspace{0.5cm}

\indent Basing on naive counting arguments, one can argue that the twist
expansion (\ref{FIT}) of the transverse moments for $Q^2 \sim \mu^2$ can
be rewritten as: $M_n^T(\mu^2) = A_n^T(\mu^2) [1 + n (\mu_n^{(4)} /
\mu)^2 - n^2 (\mu_n^{(6)} / \mu)^4]$, with $\mu_n^{(\tau)}$ approximately
independent of $n$ for $n \gsim 4$. Thus, one gets
 \be
    \label{MASS}
    \mu_n^{(4)} = \mu \sqrt{{a_n^{(4)}[T] \over n A_n^T(\mu^2)}},
    ~~~~~~~~~~~~~~~~
    \mu_n^{(6)} = \mu \left[{|a_n^{(6)}[T]| \over n^2 A_n^T(\mu^2)}
    \right]^{1/4}.
 \ee
Our results for $\mu_n^{(\tau)}$ are collected in Fig. 2, where it can
clearly be seen that the mass scales of the twist-4 and twist-6 terms of
our analysis are $\mu_n^{(4)} \simeq \mu^{(4)} \simeq 1 ~ GeV$ and
$\mu_n^{(6)} \simeq \mu^{(6)} \simeq 0.6 ~ GeV$. The value obtained for
$\mu^{(4)}$ is significantly higher than the naive expectation
$\mu^{(4)} \simeq \sqrt{<k_{\perp}^2>} \simeq 0.3 ~ GeV$ \cite{RGP77}
as well as higher than the result of other twist-4 analyses (see
\cite{JU95}).

\vspace{0.5cm}

\begin{figure}[htb]

{\epsfxsize=9cm \epsfig{file=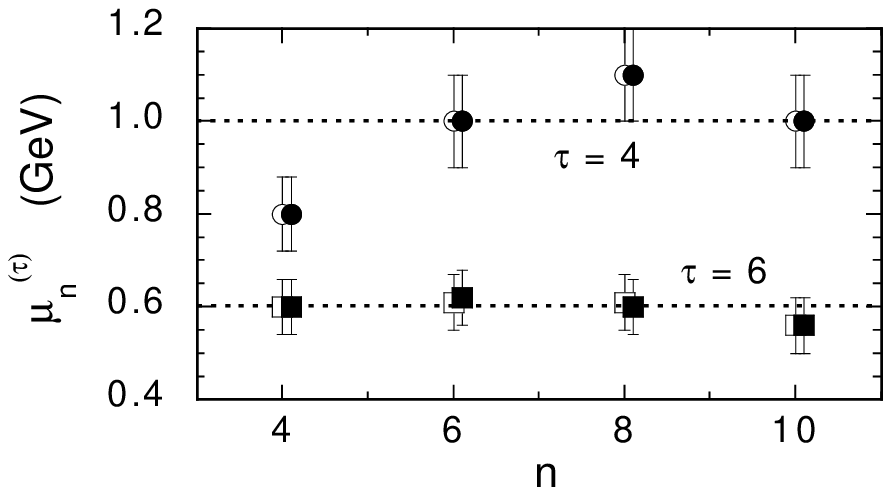}}

\vspace{-4.5cm}

\parbox{9cm}{ ~~ } \ $~~~~$ \ 
\parbox{6cm}{\small {\bf Figure 2}. The mass scale $\mu_n^{(\tau)} 
($Eq. (\ref{MASS})) of the twist-4 and twist-6 terms of our twist analysis 
(\ref{FIT}). Open and full markers correspond to the proton and deuteron 
case, while dots and squares are our results for the twist-4 and twist-6, 
respectively.} 

\end{figure}

\vspace{1.5cm}

\indent In case of the longitudinal channel the uncertainties in the
calculation of the moments are remarkably larger than those of the
transverse ones. The effects of the higher-twists are still dominant in
the second moment $M_2^L(Q^2)$ up to $Q^2$ of several $(GeV/c)^2$ (see
Fig. 3(a)). Note that the moments with $n \ge 4$ can be reproduced by
considering the leading twist plus a twist-4 term only (see Fig. 3(b)).

\vspace{0.5cm}

\begin{figure}[htb]

\centering{\epsfxsize=16cm \epsfig{file=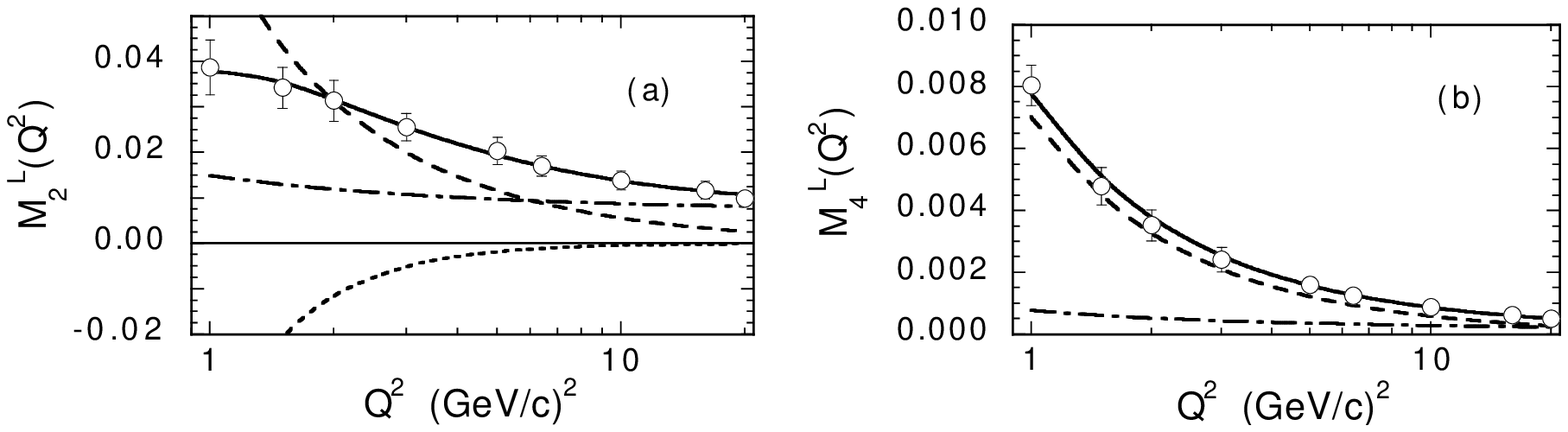}}

\vspace{0.5cm}

\parbox{0.05cm} \ $~~$ \ \parbox{15.45cm}{\small {\bf Figure 3}. The same
as in Fig. 1, but in case of the proton longitudinal structure function.}

\end{figure}

\vspace{0.5cm}

\indent The main goal of the investigation of higher-twist effects is to
disentangle the separate contributions of the various operators of a given
twist yielding the relevant multiparton correlations. This is not an easy
task, due to the contributions of many operators for any given twist (see
\cite{SV82,JS82}). Following Ref. \cite{SV82}, there are seven twist-4
operators  contributing to the second moment: three in the non-singlet
channel and four in the singlet one; the explicit expressions of these
operators can be read off from \cite{SV82}. In the non-singlet case it is
possible to write down three independent equations using, besides
$M_2^T(Q^2)$ and $M_2^L(Q^2)$, the second moment of the structure
function $F_3(x, Q^2)$, which can be determined in neutrino and
antineutrino scattering experiments, like the recent measurement
performed by the $CCFR$ collaboration \cite{CCFR} at FermiLab. The
neutrino data have been analyzed at $NNLO$ in Ref. \cite{ST97}, obtaining
a determination of the twist-4 contribution. Using all these experimental
results and adopting the notation of Ref. \cite{SV82}, we have got:
$A^{NS} = -9.0 \pm 4.5$ and $B^{NS} = -2.0 \pm 0.4$ for the quark-gluon
correlation operators, and $C^{NS} = 5.2 \pm 2.8$ for the quark-quark
correlation operator. In the singlet case the neutrino data on $F_3(x,
Q^2)$ cannot help in determining the various twist-4 operators; assuming
that the quark-quark correlation matrix element $C^S$ is the same in
neutrino and electron experiments, we have got the following constraints:
$A^S + 6 B^S = 12.5 \pm 1.8$ and $8 C^S + 5 A^S - 2 B^S = 0$.  

\indent Finally, we point out that transverse data with better quality are
still needed for $x \gsim 0.5$ and $Q^2 \lsim 10 ~ (GeV/c)^2$ as well as
more precise and systematic determinations of the $L / T$ separation are
required; for these reasons an electron facility, like  $JLab ~ @ ~ 12
~ GeV$, is a good place to investigate multiparton correlations in the
hadron structure.

\end{document}